# Evidence of universal spectral collapse at a marginal dynamical regime

Udomsilp Pinsook*, Pakin Tasee and Jakkapat Seeyangnok

Department of Physics, Faculty of Science, Chulalongkorn University, Bangkok, Thailand

*Corresponding author's email: Udomsilp.P@Chula.ac.th

**Abstract**

Incoherent electronic states in strongly correlated materials are commonly attributed to disorder or material-specific mechanisms. Here we show that incoherent spectra instead arise from self-generated dynamical disorder associated with competing fluctuations. In this regime, electron dynamics coupled to time-dependent scattering naturally produce a spectral function of the form

$$\rho(z) \propto e^{-\frac{z^2}{4}} D_\nu(z),$$

where $z$ is a scaled energy and $D_\nu$ denotes the parabolic cylinder function. This form reflects a marginal dynamical regime characterized by non-Markovian temporal correlations. Applying this scaling function to angle-resolved photoemission spectroscopy (ARPES) energy distribution curves from the cuprates $Nd_{2-x}Ce_xCuO_4$ and $Bi_2Sr_2CaCu_2O_{8+\delta}$, the Kagome metal $CsCr_3Sb_5$, and the double-layer nickelate $La_3Ni_2O_7$, we find that incoherent spectra are quantitatively described by $\rho(z)$, differing only in non-universal amplitude and energy scales. After rescaling, the datasets collapse onto a single universal curve characterized by a fixed parabolic-cylinder order $\nu = -1/2$. The observed spectral collapse indicates a fixed-point–like regime in which microscopic details such as lattice geometry, band structure, and chemical composition become irrelevant at low energies. These results establish a unified and quantitative framework for continuum-dominated ARPES spectra across diverse strongly correlated materials.

**Introduction**

Strongly correlated materials span a broad range of systems, including cuprate superconductors [1-13], nickelates [14], and Kagome compounds [15,16]. Despite pronounced differences in elemental composition, crystal structure, electronic dimensionality, and dominant interactions, these materials often display similarly unconventional electronic behavior and comparable phase diagrams [17]. In recent years, their electronic structures have been extensively investigated using angle-resolved photoemission spectroscopy (ARPES), which provides direct access to the many-body electronic spectral function, up to matrix-element effects, background contributions, and instrumental resolution. ARPES thus offers a powerful window into correlated electronic states with high energy and momentum resolution [4-16].

In ARPES, the measured intensity $I(\varepsilon,T)$ is closely related to the many-body spectral function $A(\varepsilon,k)$ multiplied by the Fermi distribution $f(\varepsilon,T)$. Identifying a universal functional form for $A(\varepsilon,k)$ that persists across distinct material classes would imply the existence of a common low-energy regime governed by scale-invariant dynamics rather than microscopic detail. Such behavior would be characteristic of a fixed-point–like structure in a phenomenological renormalization-group sense, without requiring an explicit renormalization-group construction. Similar phenomenology has been discussed in the context of marginal Fermi liquid and strange-metal behavior [18,19]. It emerged in the context of marginal Fermi liquid behavior in the cuprates, where anomalous scattering rates and continuum-dominated spectra were argued to reflect proximity to a nontrivial dynamical regime [18]. The present work differs in that it identifies an explicit universal spectral geometry governing the incoherent continuum itself.

In a conventional Fermi liquid, the spectral function may be written schematically as

$$A(\varepsilon, k) = Z_k\, \delta(\varepsilon - \varepsilon_k) + A_{\text{incoh}}(\varepsilon, k), \tag{1}$$

where the delta-function term represents coherent quasiparticle poles with residue $Z_k$, and $A_{\text{incoh}}$ denotes the incoherent continuum contribution. When quasiparticle coherence is strongly suppressed ($Z_k \to 0$), the continuum governs the spectral geometry. The central question is therefore whether this continuum is merely material-specific background or whether it obeys a universal scaling structure.

Here we present evidence that the incoherent continuum arises from a marginal dynamical regime characterized by strong suppression of long-range, and potentially long-time, electronic coherence. In this regime, the system resides near multiple competing instabilities—such as charge-density-wave order, magnetism, or superconductivity—without condensing into any single ordered phase. As a result, the low-energy electronic state is dominated by a dense continuum of fluctuating excitations rather than well-defined quasiparticles. Electronic states behave as if subject to self-generated dynamical disorder originating from competing fluctuations rather than extrinsic randomness. This framework naturally accounts for the emergence of common spectral features across different classes of strongly correlated materials.

We demonstrate that ARPES energy distribution curves (EDCs) from cuprate [8,13], nickelate [14], and Kagome [15] compounds collapse onto a single scaling function described by a parabolic cylinder form. After independent normalization of energy and intensity scales, the spectra exhibit a universal geometry characterized by a fixed order $\nu = -1/2$. This scaling behavior extends over a broad energy window and remains robust in the presence of experimental noise and weak material-specific corrections.

**Method**

Without entering into microscopic detail, a system with spatially and temporally fluctuating degrees of freedom may be described by an action of the form

$$S = S_e + S_{random} + S_{int}, \tag{2}$$

where $S_e$ represents the electronic degrees of freedom, $S_{random}$ describes a fluctuating dynamical field, and $S_{int}$ captures the coupling between electrons and this field. Despite its apparent simplicity, this structure contains essential ingredients shared by several established theoretical frameworks, including the Sachdev–Ye–Kitaev (SYK) model for strongly random systems [20,21], quantum Brownian motion descriptions of dissipative dynamics [22,23], and path-integral formulations of disorder-averaged systems [24-32].

Solutions of such models may be obtained through two complementary approaches. In the first, a canonical or unitary transformation is employed to reduce the electron–field coupling, allowing $S_{int}$ to be treated perturbatively [20,21]. In the second, the fluctuating field is integrated out, yielding an effective electronic action $S_{eff}$ that captures non-local temporal correlations [22-32]. In the present work, we adopt a solution of the latter approach.

Once the action $S$ is specified, the electronic Green's function may be obtained using a path-integral formulation,

$$G = \int \mathcal{D}x \, e^{\frac{i}{\hbar}S}. \tag{3}$$

In the absence of magnetic order and at low energies, a non-relativistic description is appropriate. In this regime, a solution of the form

$$G(t) \propto (it)^{-\frac{d}{2}} e^{-\frac{i\varepsilon^0 t}{\hbar} - \frac{\eta^2 t^2}{2}}, \tag{4}$$

naturally arises [24-25]. Here $d$ denotes the effective dimensionality, $\varepsilon^o$ is an energy shift that accounts for band alignment, and $\eta$ characterizes the strength of temporal fluctuations. The parameter $\eta$ controls the degree of coherence: in the limit $\eta \to 0$ the system approaches fully coherent quasiparticle behavior, whereas large $\eta$ corresponds to strongly incoherent dynamics. Finite $\eta$ describes a marginal dynamical regime in which electronic coherence is suppressed but not completely destroyed. Hence, the Gaussian temporal factor $e^{-\frac{\eta^2 t^2}{2}}$ reflects intrinsically non-Markovian dynamics. It also reflects the cumulative effect of dynamically fluctuating scattering fields acting on the electronic phase. This behavior contrasts with a purely Markovian process, in which memoryless scattering yields an exponential decay $e^{-\gamma t}$. The Gaussian temporal structure therefore encodes the presence of correlated dynamical fluctuations rather than simple incoherent damping. The Gaussian temporal suppression corresponds to correlated memory effects and contrasts with Markovian exponential damping, as discussed in open quantum system frameworks [33].

The corresponding spectral function is obtained via Fourier transformation,

$$\rho(\varepsilon) \propto \int dt \, G(t) \, e^{\frac{i\varepsilon t}{\hbar}}, \tag{5}$$

which yields

$$\rho(\varepsilon) = ae^{-\frac{1}{4}\left(\frac{\varepsilon^0 - \varepsilon}{\eta}\right)^2} D_{-\frac{d}{2}}\left(\frac{\varepsilon^0 - \varepsilon}{\eta}\right). \tag{6}$$

In scaling form this can be written as

$$\frac{\rho(z)}{a} = e^{-\frac{z^2}{4}} D_\nu(z), \tag{7}$$

where $a$ is an overall amplitude, $D_\nu(z)$ denotes the parabolic cylinder function, $z = \frac{\varepsilon^0 - \varepsilon}{\eta}$ is the scaled energy variable, and $\nu = -\frac{d}{2}$. For an effective one-dimensional dynamical regime (d = 1), $\nu = -\frac{1}{2}$. After appropriate normalization of energy and intensity scales, Eq. (7) yields a universal spectral form independent of microscopic material details, consistent with scale-invariant behavior characteristic of a fixed-point–like regime. This is equation is the key message of this framework.

To test the proposed universal scaling form, we assembled a cross-material set of the EDCs of ARPES selected prior to fitting based on qualitative regime criteria. Specifically, spectra were chosen that (i) are dominated by a broad continuum feature rather than sharp quasiparticle poles, (ii) exhibit a pronounced occupied-side tail and asymmetric curvature, and (iii) span a sufficiently wide energy window to evaluate the full spectral geometry. Spectra dominated by sharp coherent peaks, such as superconducting coherence features or well-defined band crossings, fall outside the regime of validity of the present framework and were therefore not considered.

In a small number of cases, minor systematic deviations from the leading scaling form are accounted for by introducing weak auxiliary contributions drawn from the same parabolic-cylinder functional family [32]. These auxiliary components do not alter the extracted exponent and represent controlled corrections to the dominant universal structure.

**Results and Discussion**

Applying the selection criteria described above, we surveyed a broad range of strongly correlated materials and identified representative systems including the cuprates $Bi_2Sr_2CaCu_2O_{8+\delta}$ (Bi2212) [8] and $Nd_{2-x}Ce_xCuO_4$ (NCCO) [13] and the double-layer nickelate $La_3Ni_2O_7$ [14] the Kagome metal $CsCr_3Sb_5$ [15]. For selected momenta $k$ at which the spectral weight is dominated by the incoherent continuum component, we approximate $A_{incoh}(\varepsilon, k) \approx \rho(\varepsilon)$. The experimentally derived spectral function is obtained from the measured ARPES intensity $I(\varepsilon, T)$ via

$$\rho(\varepsilon) = \frac{I(\varepsilon, T)}{f(\varepsilon, T)}. \tag{8}$$

No additional background subtraction or augmentation is introduced. The next step is to determine the parameters $a$, $\varepsilon^0$, and $\eta$ by fitting the original data $I(\varepsilon, T)$ to $\rho(\varepsilon) * f(\varepsilon, T)$. Auxiliary contributions are included only when necessary and take the form of small additional parabolic-cylinder components with indices $\nu = 0, 1/2, \ldots$ [32]. These terms account for minor deviations from the ideal scaling form arising from experimental resolution, background effects, or weak material-specific features, while leaving the leading universal structure unchanged (see Appendix A for fitting details). Once $a$, $\varepsilon^0$, and $\eta$ are determined, the spectra are rescaled according to $\frac{\rho(z)}{a}$. The collapsed datasets are then compared with the universal form $e^{-\frac{z^2}{4}} D_\nu(z)$, as shown in Fig. 1.

After independent normalization of energy and intensity scales, the EDCs of ARPES from cuprate [8,13], nickelate [14] and Kagome [15] compounds collapse onto a single scaling function described by the parabolic cylinder form in Eq. (7). This collapse persists over a wide energy window and across substantial differences in crystal structure, dimensionality, orbital character, and experimental conditions. The agreement is not confined to the immediate vicinity of the Fermi level but extends deep into the occupied states, where conventional quasiparticle-based descriptions typically fail. These observations provide direct evidence for a universal spectral structure governing low-energy electronic dynamics in the incoherent regime.

The scaling function is highly constrained. Once the order of the parabolic cylinder function is fixed to $\nu = -\frac{1}{2}$, only two non-universal parameters are adjusted for each dataset: an overall amplitude $a$ and the energy scaling variable $z = \frac{\varepsilon^0 - \varepsilon}{\eta}$. No additional shape parameters are introduced. The resulting fits simultaneously capture the pronounced particle–hole asymmetry [12], the extended spectral tail at negative energies, and the rapid suppression of spectral weight near the Fermi level. Importantly, these features are not reproduced simultaneously by simple Lorentzian, Gaussian, or marginal Fermi liquid line shapes, underscoring the nontrivial nature of the observed universality. In addition, we observe that $\eta$ has the same order of magnitude across different materials, see Table A1 – A4 in appendix A. Indeed, $\eta$ clusters around ~0.1 eV across our selected materials, suggesting a common incoherence scale within the marginal continuum regime. The emergence of a common incoherence scale across disparate materials echoes earlier observations of nearly universal scattering rates in strange metals [19], suggesting that the marginal continuum regime may encode a broadly shared dynamical scale.

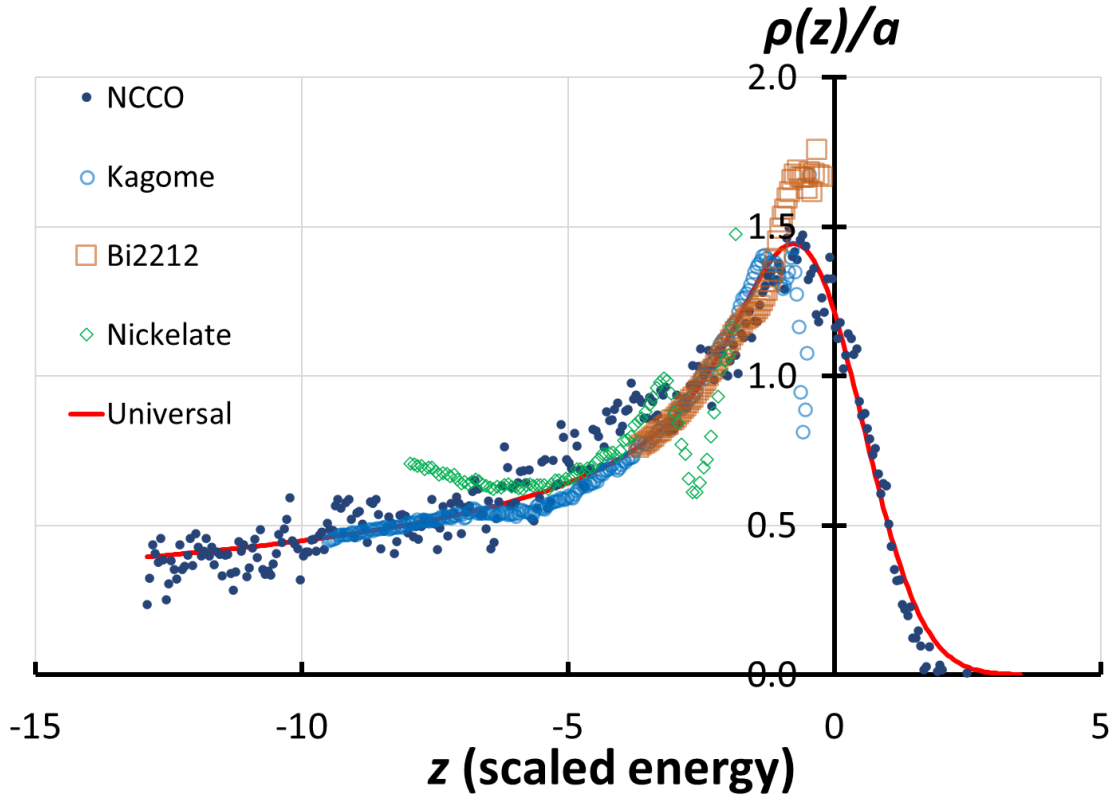

**Figure 1** shows the experimentally derived $\frac{\rho(z)}{a}$ from Bi2212 (orange squares) [8], NCCO (blue dots) [13], nickelate (green open diamonds) [14] and Kagome (open cyan circles) [15] compounds. The universal function $e^{-\frac{z^2}{4}} D_\nu(z)$ from Eq. (7) (full red line) is shown for comparison. Although, the stability of the collapse was verified by fitting $I(\varepsilon, T)$ directly via $\rho(\varepsilon) * f(\varepsilon, T)$ without performing explicit division near $\varepsilon_F$, large deviation near $z = 0$ can be noticeable because $f(\varepsilon, T) \to 0$ near $\varepsilon_F$.

From Figure A1 – A4 in appendix A, small but systematic deviations from the leading scaling form are observed in some cases, most notably in the nickelate spectra [14]. These deviations are quantitatively accounted for by introducing weak auxiliary contributions drawn from the same parabolic-cylinder functional family [32]. Such auxiliary components represent controlled corrections to the dominant scaling structure and do not alter the extracted exponent or the overall spectral collapse. Their presence is consistent with weak material-specific effects, including multi-band contributions, lattice coupling, and finite experimental resolution, while the leading universal geometry remains intact.

Notably, the electron-doped cuprate NCCO [13] exhibits excellent agreement with the leading parabolic-cylinder form over the entire energy range without requiring auxiliary corrections, despite substantial experimental noise. This robustness highlights the intrinsic nature of the scaling function: while noise obscures fine spectral features, it does not disrupt the underlying universal geometry. The ability of the same constrained functional form to describe both high-noise and high-resolution datasets strongly supports the conclusion that the scaling behavior reflects intrinsic electronic dynamics rather than fitting flexibility. In addition, the nickelate [14] represents the largest deviation from the leading scaling form among the materials considered, yet the universal component remains dominant.

The observed collapse is consistent with a marginal dynamical regime characterized by strong suppression of quasiparticle coherence and dominance of continuum spectral weight. In this regime, long-lived poles no longer control the low-energy structure. Instead, collective temporal fluctuations generate intrinsically non-Markovian dynamics that give rise to the parabolic-cylinder spectral form. In several systems, incoherent spectra appear in proximity to relatively flat dispersions, which may enhance the visibility of this regime by reducing band velocity and amplifying interaction effects. However, the scaling itself does not require a strictly flat band and is observed across materials with distinct electronic structures.

Taken together, these results indicate that the spectral collapse is a signature of a marginal dynamical regime in which long-range electronic coherence is strongly suppressed and the system resides near multiple competing instabilities. In this regime, electronic states behave as if subject to self-generated dynamical disorder arising from persistent fluctuations, leading to non-Markovian temporal correlations. The universality of the parabolic-cylinder scaling form is consistent with an effective fixed-point–like regime in a phenomenological renormalization-group sense, governing low-energy spectral properties across disparate classes of strongly correlated quantum materials.

**Conclusion**

In conclusion, we demonstrate that ARPES spectra from cuprate [8,13], nickelate [14] and Kagome [15] compounds are quantitatively described by a single universal scaling function given by the parabolic cylinder form. After independent normalization of energy and intensity scales, datasets from distinct material classes collapse onto the same spectral geometry characterized by a fixed order $v = -1/2$, with only non-universal amplitude and energy scales varying between materials. This collapse extends over a broad energy window and remains robust against experimental noise and weak material-specific corrections [32].

The observed universality indicates the presence of a marginal dynamical regime in which quasiparticle coherence is strongly suppressed and the continuum component governs the spectral structure. In this regime, electronic excitations are controlled by non-Markovian temporal dynamics arising from self-generated dynamical fluctuations rather than extrinsic disorder. Although individual systems may ultimately develop ordered phases under suitable conditions, the universal spectral form identified here reflects a common parent regime that precedes such ordering and is largely insensitive to microscopic details. The universality of these strongly correlated systems has been discussed extensively by using quantum criticality regime [17].

These results establish a unified and quantitative framework for interpreting continuum-dominated ARPES spectra across diverse strongly correlated materials. More broadly, they demonstrate that universality can emerge directly at the level of experimentally measured spectral functions, revealing that the geometry of the incoherent continuum itself encodes a fixed-point–like dynamical structure. In addition, similar incoherent continua have been reported in other strongly correlated systems, suggesting broader applicability.

## Appendix A: Fitting procedure

We fit the original data $I(\varepsilon, T)$ by using the following approximation

$$I(\varepsilon, T) \approx \{\rho(\varepsilon) + \rho_{aux}(\varepsilon)\} * f(\varepsilon, T), \tag{A1}$$

where $\rho_{aux}(\varepsilon)$ are small contributions from various orders of $D_\nu$, i.e.

$$\rho_{aux}(\varepsilon) = \sum_i a_i e^{-\frac{1}{4}\left(\frac{\varepsilon_i^0 - \varepsilon}{\eta_i}\right)^2} D_{\nu_i}\left(\frac{\varepsilon_i^0 - \varepsilon}{\eta_i}\right), \tag{A2}$$

where $\nu_i = 0, ½, 1, 3/2…$ [32]. Auxiliary terms are included only when statistically justified, and the minimal number of components necessary to remove systematic residual structure is retained. From least square fitting, we extract the parameters $a$, $\varepsilon^0$, and $\eta$ for $\rho(\varepsilon)$ (Eq. (6)). A modified variable $E = -\varepsilon$ is introduced for simplicity.

The first selected material is Nd$_{2-x}$Ce$_x$CuO$_4$ (NCCO) with x = 0.04 and T = 15 K [13]. The region in the Brillouin zone is around the in-plane momentum $k_\parallel \approx 0.7$ Å$^{-1}$ along the (0,0) – (0,π) cut. The energy range is between -0.6165 eV – 0.0455 eV, with $\mu = 0$. From the left panel of Figure A1, the raw data are taken directly from the experiment report [13] (open blue circles), compared with the fitting intensity $I(\varepsilon, T)$ (red line) with $\rho(\varepsilon)$ from Eq. (6). The error is shown by the black line in the left panel and magnified in the right panel for clarity. All fitting parameters are reported in Table A1.

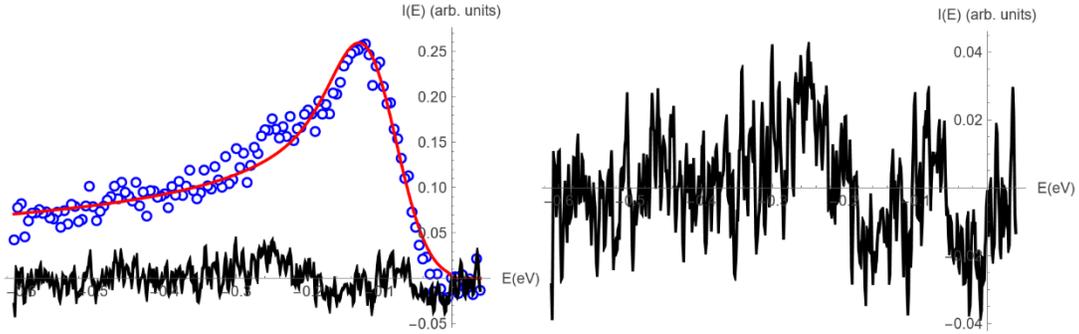

**Figure A1** (left) shows sampled experimental data [13] (opened blue circles), fitting $I(\varepsilon, T)$ (red line), and the error (black line), (right) shows the magnified error.

**Table A1** shows all the fitting parameters with their corresponding units.

| $\rho(\varepsilon)$ |  | unit |
|---|---|---|
| $\nu$ | -1/2 | order |
| $a$ | 0.180 | arb. unit |
| $\varepsilon^0$ | 0.099 | eV |
| $\eta$ | 0.040 | eV |

The second material is the cuprate Bi$_2$Sr$_2$CaCu$_2$O$_{8+\delta}$ (Bi2212) at T = 140 K [8]. The $\delta$ was chosen so that Bi2212 is overdoped with p = 0.21 and $T_c \approx 77$ K. The region in the Brillouin zone is $k_x \approx -0.8$ Å$^{-1}$ and $k_y$ are between ±0.1 Å$^{-1}$, according to the experimental setup [8]. The energy range is between -0.300 eV – 0.150 eV, with $\mu = 0$. From the left panel of Figure A2, the raw data were directly obtained from Ref [8] (open blue circles), compared with the fitting intensity $I(\varepsilon, T)$ (red line) from Eq.

(A1). The auxiliary functions, Eq. (A2), are shown by cyan and grey lines. The error is shown by the black line in the left panel and magnified in the right panel for clarity. All fitting parameters are reported in Table A2.

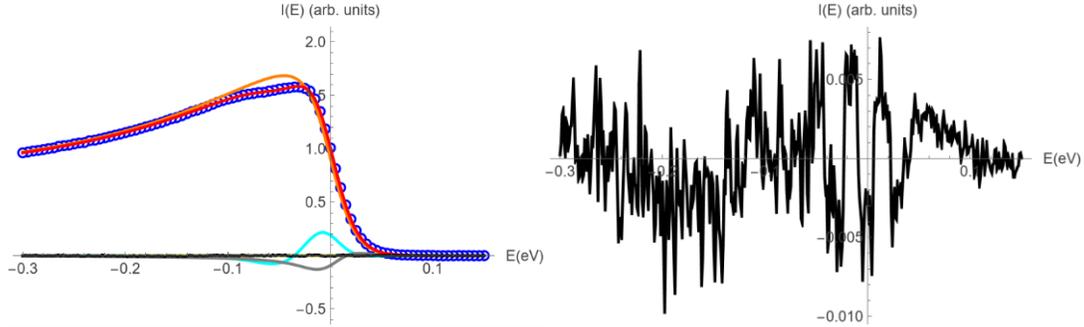

**Figure A2** (left) shows sampled experimental data [8] (opened blue circles), fitting $I(\varepsilon, T)$ (red line), and the error (black line), (right) shows the magnified error. The main contribution from $\rho(\varepsilon)$, Eq. (6), is shown by the orange line. The small auxiliary functions, Eq. (A2), are shown by cyan and grey lines.

**Table A2** shows all the fitting parameters with their corresponding units.

| $\rho(\varepsilon)$ |        | $\rho_{aux1}(\varepsilon)$ |        | $\rho_{aux2}(\varepsilon)$ |        | unit      |
|---------------------|--------|----------------------------|--------|----------------------------|--------|-----------|
| $\nu$               | -1/2   | $\nu_1$                    | 1/2    | $\nu_2$                    | 3/2    | order     |
| $a$                 | 1.266  | $a_1$                      | 0.549  | $a_2$                      | -0.475 | arb. unit |
| $\varepsilon^0$     | -0.093 | $\varepsilon_1^0$          | -0.032 | $\varepsilon_2^0$          | -0.016 | eV        |
| $\eta$              | 0.108  | $\eta_1$                   | 0.021  | $\eta_2$                   | 0.029  | eV        |

The third material is Kagome $CsCr_3Sb_5$ at T = 39.6 K [15]. The region in the Brillouin zone is at the in-plane momentum $k_\parallel = 1.0$ Å$^{-1}$ along the Γ–M cut. The energy range is between -1.1447 eV – 0.1430 eV, with $\mu = 0$. From the left panel of Figure A3, the data were provided by the authors of Ref [15] (open blue circles), compared with the fitting intensity $I(\varepsilon, T)$ (red line) from Eq. (A1). The auxiliary functions, Eq. (A2), are shown by cyan and grey lines. The error is shown by the black line in the left panel and magnified in the right panel for clarity. It is worth noting that T = 115 K gives a better fitting than the reported temperature. The effective temperature entering the Fermi function reflects experimental resolution and thermal broadening. All fitting parameters are shown in Table A3.

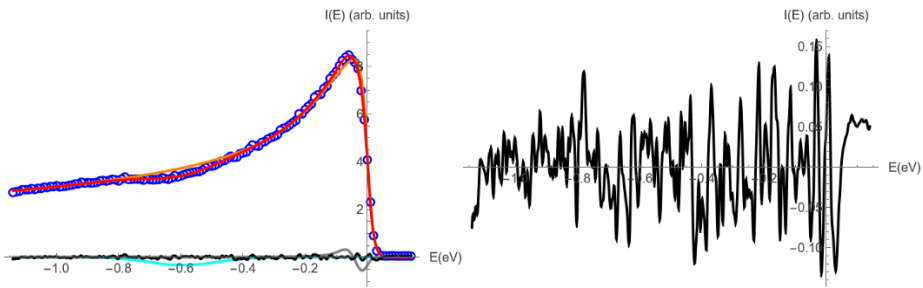

**Figure A3** (left) shows sampled experimental data [15] (opened blue circles), fitting $I(\varepsilon, T)$ (red line), and the error (black line), (right) shows the magnified error. The main contribution from $\rho(\varepsilon)$, Eq. (6), is shown by the orange line. The small auxiliary functions, Eq. (A2), are shown by cyan and grey lines.

Table A3 shows all the fitting parameters with their corresponding units.

| $\rho(\varepsilon)$ | | $\rho_{aux1}(\varepsilon)$ | | $\rho_{aux2}(\varepsilon)$ | | unit |
|---|---|---|---|---|---|---|
| $\nu$ | -1/2 | $\nu_1$ | 1/2 | $\nu_2$ | 0 | order |
| $a$ | 6.053 | $a_1$ | -0.935 | $a_2$ | -0.330 | arb. unit |
| $\varepsilon^0$ | -0.112 | $\varepsilon_1^0$ | 0.024 | $\varepsilon_2^0$ | 0.592 | eV |
| $\eta$ | 0.132 | $\eta_1$ | 0.028 | $\eta_2$ | 0.108 | eV |

Finally, the last material is the double-layer nickelate La$_3$Ni$_2$O$_7$ at T = 163 K [14]. The region in the Brillouin zone is around (π,π). The energy range is between -0.3972 eV – 0.0624 eV, with $\mu = 0$. From the left panel of Figure A4, the data were extracted from the experiment report [14] (open blue circles), compared with the fitting intensity $I(\varepsilon, T)$ (red line) from Eq. (A1). The auxiliary functions, Eq. (A2), are shown by cyan and grey lines. They are quite subdominant but non-negligible corrections, see the left panel of Figure A4. They show that nickelate has structured deviation from that of $\rho(\varepsilon)$ from Eq. (6). However, the universal component (orange line) still dominates. The error is shown by the black line in the left panel and magnified in the right panel for clarity. The error is in the window of ± 0.25 and mostly structureless except minor features. In the nickelate band structure, there exists a real pole below -0.3 eV. This explains the discrepancy at lower energy. All fitting parameters are reported in Table A4.

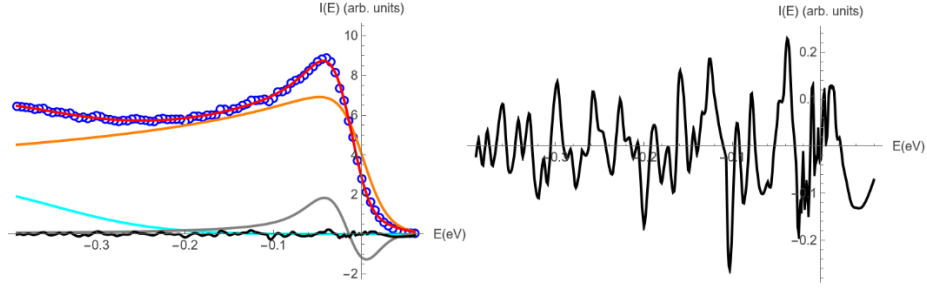

**Figure A4** (left) shows sampled experimental data [14] (opened blue circles) and fitting $I(\varepsilon, T)$ (red line), and the error (black line), (right) shows the magnified error. The main contribution from $\rho(\varepsilon)$, Eq. (6), is shown by the orange line. The auxiliary functions, Eq. (A2), are shown by cyan and grey lines. It is worth noting that the magnitudes of the auxiliary functions are quite noticeable in this case.

Table A4 shows all the fitting parameters with their corresponding units.

| $\rho(\varepsilon)$ | | $\rho_{aux1}(\varepsilon)$ | | $\rho_{aux2}(\varepsilon)$ | | unit |
|---|---|---|---|---|---|---|
| $\nu$ | -1/2 | $\nu_1$ | 1/2 | $\nu_2$ | 0 | order |
| $a$ | 7.239 | $a_1$ | -5.549 | $a_2$ | 2.333 | arb. unit |
| $\varepsilon^0$ | -0.247 | $\varepsilon_1^0$ | -0.006 | $\varepsilon_2^0$ | 0.466 | eV |
| $\eta$ | 0.120 | $\eta_1$ | 0.027 | $\eta_2$ | 0.118 | eV |

**Acknowledgement**

PT is funded by the 90th Anniversary of Chulalongkorn University Scholarship under the Ratchadapisek Somphot Endowment Fund, and JS is funded by the Second Century Fund (C2F), Chulalongkorn University. We would like to express our sincere gratitude to Prof. Ming Yi, Prof. Di-Jing Huang and Dr.

Yucheng Guo for providing the raw ARPES data of Kagome CsCr$_3$Sb$_5$. We are deeply grateful for their kindness.

**AI declaration**

The authors acknowledge the use of ChatGPT (OpenAI) for assistance in editing, language refinement, aimed for improving clarity and presentation of this manuscript.
**References**

[1] J.G. Bednorz and K.A. Müller, "Possible High Tc Superconductivity in the Ba-La-Cu-O System", Z. Phys. B Cond. Matter 64, 189 (1986).

[2] M. K. Wu, J. R. Ashburn, C. J. Torng, P. H. Hor, R. L. Meng, L. Gao, Z. J. Huang, Y. Q. Wang, and C. W. Chu, "Superconductivity at 93 K in a new mixed-phase Y-Ba-Cu-O compound system at ambient pressure", Physical Review Letters 58, 908 (1987).

[3] H. Maeda; Y. Tanaka; M. Fukutumi, and T. Asano, "A New High-Tc Oxide Superconductor without a Rare Earth Element", Jpn. J. Appl. Phys. 27 (2), L209 – L210 (1988).

[4] I. M. Vishik, M. Hashimoto, R.-H. He, W.-S. Lee, F. Schmitt, D. Lu, R. G. Moore, C. Zhang, W. Meevasana, T. Sasagawa, S. Uchida, K. Fujita, S. Ishida, M. Ishikado, Y. Yoshida, H. Eisaki, Z. Hussain, T. P. Devereaux, and Z.-X. Shen, "Phase competition in trisected superconducting dome", PNAS 109 (45), 18332-18337 (2012).

[5] H. Anzai, A. Ino, M. Arita, H. Namatame, M. Taniguchi, M. Ishikado, K. Fujita, S. Ishida and S. Uchida, "Relation between the nodal and anti-nodal gap and critical temperature in superconducting Bi2212", Nature Communications 4, 1815 (2013).

[6] M. Hashimoto, I. M. Vishik, R.-H. He, T. P. Devereaux and Z.-X. Shen, "Energy gaps in high-transition-temperature cuprate superconductors", Nature Physics 10, 483-495 (2014).

[7] A. A. Kordyuk, "Pseudogap from ARPES experiment: three gaps in cuprates and topological superconductivity", Low Temperature Physics 41, 319–341 (2015).

[8] S. D. Chen, M. Hashimoto, Y. He, D. Song, J. F. He, Y. F. Li, S. Ishida, H. Eisaki, J. Zaanen, T. P. Devereaux, D. H. Lee, D. H. Lu, and Z. X. Shen, "Unconventional spectral signature of Tc in a pure d-wave superconductor", Nature 601, 562–567 (2022).

[9] E. G. Maksimov, M. L. Kulic, and O. V. Dolgov, "Bosonic Spectral Function and the Electron-Phonon Interaction in HTSC Cuprates", Advances in Condensed Matter Physics 2010, 423725 (2010).

[10] A. Kordyuk, S. V. Borisenko, T. K. Kim, K. Nenkov, M. Knupfer, J. Fink, M. S. Golden, H. Berger, R. Follath, "Origin of the peak-dip-hump structure in the photoemission spectra of Bi$_2$Sr$_2$CaCu$_2$O$_8$", Physical Review Letters 89, 077003 (2002).

[11] T. L. Miller, W. Zhang, J. Ma, H. Eisaki, J. E. Moore, and A. Lanzara, "Interplay of superconductivity and bosonic coupling in the peak-dip-hump structure of Bi$_2$Sr$_2$CaCu$_2$O$_{8+\delta}$", Physical Review B 97, 134517 (2018).